\documentstyle[preprint,aps]{revtex}
\tightenlines

\begin{document}

\title{Canonical Transformation of the Three-Band Hubbard
Model and Hole  Pairing }
\author{Michele Cini, Gianluca Stefanucci and Adalberto Balzarotti}
\address{INFM, Dipartimento di Fisica,Universit\`{a} di Roma Tor Vergata, Via\\
della Ricerca Scientifica 1- 00133 Roma, Italy}
\date{\today}
\maketitle

\begin{abstract}

We propose a canonical tranformation approach to the effective interaction $W_{eff}$
between two holes, based on the three-band Hubbard model but ready to include
extra interactions as well.  An effective two-body Hamiltonian can in principle be
obtained including any kind of virtual intermediate states. We derive the closed-form
analytic expression of the effective interaction including 4-body virtual states,
describing the exchange of an electron-hole pair to all orders. The resulting integral
equation, valid for the full plane, leads to a Cooper-like instability of the Fermi
liquid. The two-hole bound states generalize those reported earlier in cluster
calculations by exact diagonalisation methods.

\end{abstract}


\noindent

\narrowtext


The three-band Hubbard model Hamiltonian has often been used to characterize the
electronic properties of high-$T_{C}$ superconductors\cite{kn:bm} as observed in
electron spectroscopy experiments.  The model is: 
\begin{equation}
H=H_{0}+W  \label{h0}
\end{equation}
\noindent
where the independent hole hamiltonian reads, in the site representation

\begin{equation}
H_{0}={\sum_{Cu}}\varepsilon _{d}n_{d}+{\sum_{O}}\varepsilon _{p}n_{p}+{\
t\sum_{n.n.}}\left[ c_{p}^{+}c_{d}+h.c.\right]  \label{1}
\end{equation}
\noindent
where n.n. stands for nearest neighbors. The on-site repulsion Hamiltonian
will be denoted by 
\begin{equation}
W={\sum_{i}}U_{i}n_{i+}n_{i-},  \label{w}
\end{equation}

\noindent
where $U_{i}=U_{d}$ for a Cu site, $U_{i}=U_{p}$ for an Oxygen. We also
considered first-neighbor O-O hopping and off-site interactions\cite{kn:cb1}, but they
were dropped when it was clear that they were not really important. The point
symmetry Group of the Cu-O plane is $C_{4v}$, and its characters are listed in Table
I. We wish to show that holes are paired in a Cooper-like fashion in the
ground state of this popular model. Preliminarly, however, we must introduce the
W=0 pairs. Consider a determinantal 2-hole eigenstate of $H_{0}$ built from
degenerate Bloch states. Can we superpose such wave functions to obtain simultaneous
eigenstates  of $H$,$H_{0}$ and $W$? For a general system, of course, we can't, but
the symmetry of the problem enables us to achieve the result, and in addition to
choose the eigenvalue W=0 of the on-site interaction. Indeed, omitting the band
indices,let 
\begin{equation}
d[k]=\left\| k_{+},-k_{-}\right\| =c_{k,+}^{\dagger}c_{-k,-}^{\dagger}|vac> 
\label{3}
\end{equation}
 be a two-hole determinantal state derived from the Bloch 
eigenfunctions of $H_{0}$. We introduce the determinants $Rd[k]=d[Rk]=d[k_{R}],R\in
C_{4v}$ , and the projected states 
\begin{equation}
\Phi _{\eta }\left[ k\right] =\frac{1}{\sqrt{8}}{\sum_{R\in C_{4v}}}\chi
^{\left( \eta \right) }\left( R\right) Rd[k]  \label{4}
\end{equation}
\noindent
where $\chi ^{\left( \eta \right) }(R)$ is the character of the operation $R$
in the Irreducible Representation (Irrep) $\eta $. In the non-degenerate
Irreps, the operations that produce opposite $k_{R}$ have the same character,
and the corresponding projections lead to singlets. The explicit
W=0 singlet pair states are: 
\begin{eqnarray}
\Phi _{^{1}B_{2}}\left[ k,r_{1},r_{2}\right] =\frac{\chi _{0}}{\sqrt{2}}
\{
\cos\left[ k(r_{1}-r_{2})\right] \phi \left( k,r_{1}\right) \phi \left(
k,r_{2}\right)   \nonumber \\
-\cos\left[ k_{C_{4}}(r_{1}-r_{2})\right] \phi \left( k_{C_{4}},r_{1}\right) \phi
\left( k_{C_{4}},r_{2}\right)\nonumber \\
 -\cos\left[ k_{\sigma}(r_{1}-r_{2})\right] \phi
\left( k_{\sigma },r_{1}\right) \phi \left( k_{\sigma },r_{2}\right)
 \nonumber \\
+\cos\left[ k_{\sigma^{\prime} }(r_{1}-r_{2})\right] \phi
\left( k_{\sigma^{\prime} },r_{1}\right) \phi \left( k_{\sigma^{\prime}
},r_{2}\right)
\}\label{b2e}
\end{eqnarray}

and  
\begin{eqnarray}
\Phi _{^{1}A_{2}}\left[ k,r_{1},r_{2}\right]  =\frac{\chi _{0}}{\sqrt{2}}
\{
\cos\left[ k(r_{1}-r_{2})\right] \phi \left( k,r_{1}\right) \phi \left(
k,r_{2}\right)   \nonumber \\
+\cos\left[ k_{C_{4}}(r_{1}-r_{2})\right] \phi \left( k_{C_{4}},r_{1}\right) \phi
\left( k_{C_{4}},r_{2}\right)\nonumber \\
 -\cos\left[ k_{\sigma}(r_{1}-r_{2})\right] \phi
\left( k_{\sigma },r_{1}\right) \phi \left( k_{\sigma },r_{2}\right)
 \nonumber \\
-\cos\left[ k_{\sigma^{\prime} }(r_{1}-r_{2})\right] \phi
\left( k_{\sigma^{\prime} },r_{1}\right) \phi \left( k_{\sigma^{\prime}
},r_{2}\right)
\},\label{a2e}
\end{eqnarray}
\noindent
where $\chi_{0}$ is a singlet spin function and $\phi$ is the periodic part of
the Bloch function. Using the explicit Bloch states we can verify that both vanish
for $r_{1}=r_{2}$ . Therefore they have the W=0 property.
Turning now to the many-body problem, suppose the Cu-O plane is in its ground
state with chemical potential $\mu\equiv E_{F}$ and a couple of extra holes are
added. We can consider the W=0 pairs as bare quasiparticles for which no repulsion
barrier needs to be overcome for pairing. Do the dressed quasiparticles look like
Cooper pairs?

In Cooper theory\cite{kn:coop}, an effective interaction involving phonons
is introduced via an approximate canonical transformation. Our problem is
similar, except that the holes can exchange electron-hole pairs instead of
phonons. Now the vacuum is the filled Fermi {\em sphere} and W connects the
determinants $d[k]$ to the 4-body (3 hole-1 electron) determinants  that carry no
quasi-momentum; they are of the form

\begin{equation}
|\alpha >=|\left\| \left( k^{\prime }+q+k_{2}\right) _{+},\bar{k}%
_{2-},-q_{-},-k_{-}^{\prime }\right\| >  \label{alfas}
\end{equation}
\noindent
where $\bar{k}_{2}$ is the electron state and pedices refer to the spin
direction; those with opposite spin indices contribute similarly and yield a
factor of 2 at the end. These are eigenstates of $H_{0}$:
\begin{equation}
H_{0}\left| \alpha \right\rangle =E_{\alpha }\left| \alpha \right\rangle 
\end{equation}

The interaction matrix element reads:

\begin{eqnarray}
<\left\| \left( k^{\prime }+q+k_{2}\right) _{+},\bar{k}
_{2-},-q_{-},-k_{-}^{\prime }\right\| |W|d[s]>=  \nonumber \\
\delta \left( q-s\right) U\left( q+k^{\prime }+k_{2},-k^{\prime
},s,k_{2}\right)  \nonumber \\
-\delta \left( k^{\prime }-s\right) U\left( q+k^{\prime
}+k_{2},-q,s,k_{2}\right)  \label{me}
\end{eqnarray}
\noindent
where the $U$ matrix is computed on the Bloch basis. If we want to keep a close
analogy with the Cooper theory, we temporarily consider $W$ as a small perturbation
and look for an approximate canonical transformation such that the new Hamiltonian

\begin{equation}
\tilde{H}=e^{-\Lambda}He^{\Lambda} \label{cano}
\end{equation} 
\noindent
contains no first-order terms. Here, $ \tilde{H}$ operates on the space of pairs,
since the 3 holes-1 electron intermediate states have been decoupled by the canonical
transformation. This is accomplished if

\begin{equation}
W+\left[H_{0}, \Lambda\right]=0  \label{hs2}
\end{equation}

because then, to second order,
\begin{equation}
\tilde{H}=H_{0}+\frac{1}{2}\left[W,\Lambda\right].\label{hs3}
\end{equation}

  Assuming that the denominators do not vanish (more about that later)  we obtain

\begin{equation}
\left\langle \alpha \right| \Lambda\left| s\right\rangle =\frac{\left\langle
\alpha \right| W\left| s\right\rangle }{E_{s}-E_{\alpha }}
\end{equation}

We may write
\begin{equation}
\tilde{H}=H_{0}+F+\tilde{W}_{eff}\label{hs4}
\end{equation}
\noindent
where $F$ is diagonal in the pair space, like $H_{0}$, and corresponds to the
unlinked self-energy diagrams, while the effective interaction operator is
$ \tilde{W}_{eff} $. Like in Cooper theory, F will be dropped. We obtain

\begin{eqnarray}
2\left\langle p \right| F+\tilde{W}_{eff}\left| s\right\rangle
\nonumber\\
=\sum_{\alpha}W_{p,\alpha}W_{\alpha,s}\left[
\frac{1 }{E_{p}-E_{\alpha }}+\frac{1 }{E_{s}-E_{\alpha }}
\right]\label{u}
\end{eqnarray}

Using the interaction matrix element (\ref{me}), the product in (\ref{u}) yields 4
terms;  two  are proportional to $\delta\left(p-s\right)$ and belong to $F$, while
the cross terms yield identical contributions to the effective interaction. 

After long algebra, we write the interaction between symmetry projected states (with
$s$ and $p$ empty):

\begin{equation}
\left\langle \Phi_{\alpha}\left[ p\right] \right| \tilde{W}_{eff}\left|
\Phi_{\alpha}\left[ s\right] \right\rangle =
\sum_{R}\chi^{\left(\alpha\right)}\left(R\right)
 \left\langle d\left[ p\right] \right| \tilde{W}_{eff}\left|
Rd\left[ s\right] \right\rangle
\label{wfi}
\end{equation}
where, explicitly,
\begin{eqnarray}
\left\langle d\left[ p\right] \right| \tilde{W}_{eff}\left| d\left[
s\right] \right\rangle =2
\sum_{k}^{occ}\Theta \left( \varepsilon \left( s+p+k\right) -E_{F}\right)
\nonumber\\
\times U\left( s+p+k,-p,s,k\right) U\left( p,k,s+p+k,-s\right)
 \nonumber\\
\times [\frac{1}{\varepsilon \left( s+p+k\right) -\epsilon \left( k\right)
-\epsilon \left( s\right) +\epsilon \left( p\right) }
\nonumber\\
+\frac{1}{\varepsilon \left( s+p+k\right) -\epsilon \left( k\right)
-\epsilon \left( p\right) +\epsilon \left( s\right) }] \label{wd}
\end{eqnarray}

In the continuum limit, the integral must be understood as a principal part,
 because the denominators vanish in a domain of zero measure within the
integration domain. However, this complication is an artifact of
the perturbation approach, as will soon become clear.
The approximate canonical transformation is open to question
because it assumes that $W$ be weak. Therefore, it is important
to put the theory on a clearer and firmer basis, by removing the coupling to the
$\alpha$ states to all orders.

The transformation  (\ref{cano},\ref{hs2}) corresponds to a second-order diagram for the 
two-hole amplitude, and if one of the hole lines is closed on itself 
one gets the corresponding self-energy. The expansion of $\Lambda$
can be continued systematically to produce the perturbation series. Including all the diagrams of a 
generalized RPA would lead to something like the well-known FLEX approximation 
\cite{bsw} whose implications for the superconductivity in 
the three-band Hubbard model have been explored recently in a series 
of papers\cite{eb}. A related self-consistent and 
conserving T-matrix approximation has been proposed by Dahm and 
Tewordt\cite{dt} for the excitation spectra in the 2D Hubbard model; 
we  mention incidentally that recently diagrammatic methods have been 
successfully applied to the photoelectron spectra of the Cuprates in other contexts 
too, like the spin-fermion model\cite{sps}.

Realizing the key r\^{o}le of symmetry and W=0 pairs in this problem enhances intuition
besides simplifying the perturbation formalism considerably. Here, 
we wish to take advantage of the W=0 pairs to propose a non-perturbative  approach to 
pairing based on a different way of performing the canonical transformation.
We write the ground state wave function with two
added holes as a superposition of two-body states (roman indices) and 4-body ones:

\begin{equation}
|\Psi _{0}>={\sum_{m}}a_{m}|m>+{\sum_{\alpha }}b_{\alpha }|\alpha
>. \label{psi00}
\end{equation}

 Schr\"{o}dinger's equation then yields
\begin{equation}
\left( E_{m}-E_{0}\right) a_{m}+{\sum_{m^{\prime }}}a_{m^{\prime
}}V_{m,m^{\prime }}+{\sum_{\alpha }}b_{\alpha }W_{m,\alpha }=0  \label{sys1}
\end{equation}

\begin{equation}
\left( E_{\alpha }-E_{0}\right) b_{\alpha }+{\sum_{m^{\prime }}}a_{m^{\prime
}}W_{\alpha ,m^{\prime }}=0  \label{sys2}
\end{equation}
where $E_{0}$ is the ground state energy. $V_{m^{\prime },m}$ vanishes for W=0 pairs in our
model; however we keep it for generality, since it allows to introduce the effect of
phonons\cite{kn:pietro,kn:iad} and any other indirect
interaction\cite{kn:bob,kn:pw,kn:sus} that we are not considering. Solving for
$b_{\alpha }$ and substituting in (\ref{sys1}) we exactly decouple the 4-body states.
The eigenvalue problem is now 
\begin{equation}
\left( E_{0}-E_{m}\right) a_{m}=\sum_{m^{\prime}}  a_{m^{\prime
}}\left\{V_{m,m^{\prime }}+\left\langle m|S[E_{0}]|m^{\prime }\right\rangle \right\}
,  \label{dec}
\end{equation}
where
\begin{equation}
\left\langle m|S\left[ E_{0}\right] |m^{\prime }\right\rangle ={\sum_{\alpha
}}\frac{<m|W|\alpha ><\alpha |W|m^{\prime }>}{E_{0}-E_{\alpha }}.  \label{wt1}
\end{equation}
This is of the form of a Schr\"{o}dinger equation with eigenvalue
$E_{0}$ for pairs with an effective interaction $V+S$. Then we interpret
$a_{m}$ as the wave function of the dressed pair, which is acted upon by an
effective hamiltonian $\tilde{H}$. This is the canonical transformation we were
looking for. However, the scattering operator $S$ is of the form $S=W_{eff}+F,$
where $W_{eff}$ is the effective interaction between dressed holes, while $F$ is
a forward scattering operator, diagonal in the pair indices $m$ ,$m^{\prime }$
which accounts for the self-energy corrections of the one-body propagators: it is
evident that it just redefines the dispersion law $E_{m}$, and,
essentially, renormalizes the chemical potential. Therefore $F$ must be
dropped, as in Cooper theory\cite{kn:coop} and above. Therefore the effective
Schr\"{o}dinger equation for the pair reads

\begin{equation}
\left( H_{0}+V+W_{eff}\right) |a>=E_{0}|a>  \label{cang}
\end{equation}
and we are interested in the possibility that $E_{0}=2E_{F}-\left| \Delta \right|
$, with a positive binding energy $\left| \Delta \right| $ of the
pair. The $V$ interaction just adds to $W_{eff}$, and this feature allows to
include in our model the effects of other pairing mechanisms, like off-site
interactions, inter-planar coupling and phonons. Again, the product in the numerator of
(\ref{wt1})  yields 4 terms; two are proportional to $\delta (p-s)$ and belong to F, while
the cross terms yield identical contributions to $W_{eff}$. Hence we obtain the effective
interaction between W=0 pairs:

\begin{eqnarray}
\left\langle \Phi _{\eta }\left[ p\right] \right| W_{eff}\left| \Phi _{\eta
}\left[ s\right] \right\rangle =4{\sum_{R\in C_{4v}}}\chi
^{\left( \eta \right) }\left( R\right) {\sum_{k}^{occ}}
\Theta \left( \varepsilon \left( Rs+p+k\right) -E_{F}\right)
\nonumber\\
\times  \frac{U\left( Rs+p+k,-p,Rs,k\right)U\left( p,k,Rs+p+k,-Rs\right)}
{\varepsilon \left( Rs+p+k\right) -\epsilon \left( k\right)
+\epsilon \left( s\right) +\epsilon \left( p\right)-E_{0} }
 \label{weffective}
\end{eqnarray}
The sum is over occupied $k$ with empty $Rs+p+k$. There are no vanishing
denominators in the new expression, if $E_{0}$ is off the continuum. The $p$ and
$s$ indices run over 1/8 of the Brillouin Zone. We denote such a set of empty
states $e/8$, and cast the result in the form of a (Cooper-like) Schr\"{o}dinger equation

\begin{equation}
2\varepsilon \left( k\right) a\left( k\right) +\stackrel{e/8}{
\sum_{k^{\prime }}}W_{eff}\left( k,k^{\prime }\right) a\left( k^{\prime
}\right) =E_{0}a\left( k\right)  \label{inteq}
\end{equation}
for a self-consistent calculation of $E_{0}$ (since $W_{eff}$ depends on the
solution). Let $N_{C}$ be the number of cells in the crystal. The $U$ matrix
elements scale as $N_{C}^{-1}$ and therefore $W_{eff}$ scales in the same
way. For an infinite system, $N_{C}\rightarrow \infty $ , this is a well
defined integral equation.
In principle, we can do better. By a canonical transformation one can obtain
an effective Hamiltonian which describes the propagation of a pair of
\underline{dressed} holes, and includes all many-body effects. The exact
many-body ground state with two added holes may be expanded in terms of
excitations over the vacuum (the non-interacting Fermi {\em sphere}) by a
configuration interaction: 
\begin{equation}
|\Psi _{0}>={\sum_{m}}a_{m}|m>+{\sum_{\alpha }}b_{\alpha }|\alpha >+{\
\sum_{\beta }}c_{\beta }|\beta >+....  \label{psi0}
\end{equation}
here m runs over pair states, $\alpha $ over 4-body states ($2$ holes and $1$
e-h pair), $\beta $ over 6-body ones ($2$ holes and $2$ e-h pairs), and so on.
To set up the Schr\"{o}dinger equation, we consider the effects of the
operators on the terms of $|\Psi _{0}>$. We write:

\begin{equation}
H_{0}|m>=E_{m}|m>,\;H_{0}|\alpha >=E_{\alpha }|\alpha >,...  \label{h0m}
\end{equation}
and since W can create or destroy up to 2 e-h pairs,

\begin{eqnarray}
W|m>={\sum_{m^{\prime }}}V_{m^{\prime },m}|m^{\prime }>+{\sum_{\alpha }}%
|\alpha >W_{\alpha ,m}  \nonumber \\
+{\ \sum_{\beta }}|\beta >W_{\beta ,m},  \label{wm}
\end{eqnarray}
and the like. The Schr\"{o}dinger equation now gives an infinite system for the amplitudes
of 2n-body states; however we can show \cite{kn:sub} that these amplitudes can be
successively decoupled, producing a renormalization of W matrix elements and energy
eigenvalues $E_{\alpha},E_{\beta}$ and so on. In principle, the method applies to all the
higher order interactions, and we can recast our problem as if only 2 and 4-body
states existed. If one calculates  $W_{eff}$ neglecting 6-body and higher excitations, at
least the structure of the solution is exact when expressed in terms of renormalized
matrix elements.

The Hubbard model, with V=0, leads to pairing, and the mechanism was first
discovered in cluster studies\cite{kn:cb1}. Now we can better understand those results,
since the above theory applies not only to the plane but also to clusters, provided that
they are fully symmetric and allow W=0 pairs. The symmetry of the cluster is essential,
because only fully symmetric clusters allow such solutions. The planar lattice
structure is also essential, because no W=0 pairs occur in 3D or in a continuous
model. We studied\cite{kn:cb1,kn:cb2,kn:cb3,kn:cb4} the fully symmetric clusters with up to
21 atoms by exact diagonalisation. The main difference is that in the clusters
the symmetries of W=0 pairs were found\cite{kn:cb5,kn:cbs} to be $^{1}B_{2}$ and
$^{1}A_{1}$. The reason for having $^{1}A_{1}$ instead of $^{1}A_{2}$ is a twofold size
effect. On one hand,$^{1}A_{1}$ pairs have the W=0 property only in the small clusters,
having the topology of a cross, and belonging to the $S_{4}$ Group, but do not generalize as
such to the full plane, when the symmetry is lowered to $C_{4v}$; on the other
hand, those small clusters admit no solutions of $^{1}A_{2}$ symmetry.
To monitor pairing, initially we used\cite{kn:cb1} a definition of $\Delta$ taken from
earlier cluster calculations\cite{kn:hirsch,kn:bal} (where, however, only unsymmetric
clusters were considered, and this mechanism could not operate):

\begin{equation}
\Delta=E(N+2)+E(N)-2E(N+1),  \label{delta}
\end{equation}
where $E(N)$ is the ground state energy of the cluster with N holes, as
obtained by exact diagonalisation. Pairing, that is, $\Delta<0$, was found 
when (and only when) the least bound holes formed a W=0 pair. In small clusters this
conditions the occupation number, because the holes must partly occupy degenerate (x,y)
orbitals, while in the full plane the W=0 pairs (\ref{b2e},\ref{a2e}) exist at the
Fermi level for any filling. The second-order approximation
$\Delta^{(2)}$  was obtained by ground state energy diagrams (modified for
degenerate ground states, when appropriate); the resulting expression\cite{kn:cb5}
is clearly just the second-order approximation to the diagonal terms of
(\ref{weffective}). Then, we computed to second-order the two-hole amplitude for
holes of opposite spins in the degenerate (x,y) orbitals.  We demonstrated that this
produced an effective interaction, which pushes down the singlet and up the triplet
by
$\left| \Delta ^{\left( 2\right) }\right| $.   Good agreement between the second-order
calculation and the numerical exact diagonalisation results supported the interpretation.
Thus, we have shown that in the symmetric Cu-O clusters genuine pairing takes place,
due to an effective interaction which is attractive for singlets and repulsive for
triplets.   The cluster calculations\cite {kn:cb2,kn:cb3,kn:cb4} showed that W=0 pairs
are the $^{\prime\prime }bare^{\prime \prime }$ quasiparticles that, when
$^{\prime\prime }dressed^{\prime \prime }$, become a bound state. 
That approach is inherently limited by the small size of solvable clusters, but
allows a very explicit display of paired hole properties, that even show
superconducting flux-quantization\cite{kn:cb5,kn:cbs}.

The  equations (\ref{weffective},\ref{inteq})  allow to extend the study of pairs to
the full plane. The integrand is very discontinuous because of Umklapp terms and of
the limitations to occupied or empty states; moreover, its angular dependences are
involved. 
 The numerical solution is far from trivial, and the method that we developed will
be presented elsewhere, with the numerical results. The main finding is the 
instability of the Fermi liquid in the model at hand.  For both $^{1}B_{2}$ and
$^{1}A_{2}$ we find pairing with a doping-dependent $\Delta$. 
The three-band Hubbard model might be too idealized to allow a detailed
comparison with experiments; however we stress that contributions from phonons and other
 mechanisms can be included by a non zero $V$. The approach presented here is more general
than the model we are using, and can be applied to realistic Hamiltonians.

This work has been supported by the Is\-ti\-tu\-to Na\-zio\-na\-le di Fisica
della Materia. We gratefully acknowledge A. Sa\-gnot\-ti, Universit\`{a} di
Roma Tor Ver\-ga\-ta, for useful and stimulating discussions.



\smallskip


\begin{table}[tbp]
\begin{center}
\begin{tabular}{lclclclclclcl}
C$_{4v}$ & E & C$_{2}$ & 2C$_{4}$ & 2$\sigma $ & 2$\sigma ^{\prime }$ &  & 
&  &  &  &  &  \\ 
A$_{1}$ & 1 & 1 & 1 & 1 & 1 &  &  &  &  &  &  &  \\ 
A$_{2}$ & 1 & 1 & 1 & -1 & -1 & $R_{z}$ &  &  &  &  &  &  \\ 
B$_{1}$ & 1 & 1 & -1 & 1 & -1 & $x^{2}-y^{2}$ &  &  &  &  &  &  \\ 
B$_{2}$ & 1 & 1 & -1 & -1 & 1 & $xy$ &  &  &  &  &  &  \\ 
E & 2 & -2 & 0 & 0 & 0 & $\left( x,y\right) $ &  &  &  &  &  & 
\end{tabular}
\end{center}
\caption{The Character Table of the $C_{4v}$ Group}
\label{Table I}
\end{table}

\bigskip

\end{document}